# FUTURE PROOF FOR PHYSICS:
## Preserving the Record of SLAC

Jean Marie Deken

*Stanford Linear Accelerator Center*
*Stanford University*
*2575 Sand Hill Road MS82*
*Menlo Park CA 94025*

**Abstract**
Paper provides a brief introduction to SLAC, discusses the origins of the SLAC Archives and History Office, its present-day operations, and the present and future challenges it faces in attempting to preserve an accurate historical record of SLAC's activities.

Work supported by Department of Energy contract DE-AC03-76SF00515

**What is SLAC?**

The idea for a two-mile linear accelerator at Stanford University was conceived in 1956, proposed in 1957, and authorized by the United States (US) Congress in 1961. Initially called "Project M," the venture was renamed "The Stanford Linear Accelerator Center" (SLAC) in August of 1960. The original contract between Stanford University and the U.S. Atomic Energy Commission was signed on April 30, 1962: construction began the following July and was completed February 10, 1966. SLAC's official dedication occurred on September 9, 1967.[1] SLAC is owned by the United States government, and is operated for the US Department of Energy by Stanford University. Its present-day mission is to design, construct and operate state-of-the-art electron accelerators and related experimental facilities for use in high-energy physics and synchrotron radiation research.

SLAC occupies 430 acres of the Stanford University campus near the intersection of Sand Hill Road and US Highway 280 in northern California. In fiscal year 2002, its budget was $209 million: it employed a staff of 1,467 (full-time equivalents); and hosted 3,000 users from a variety of institutions, including universities (147), industry (46), government laboratories (30), and foreign countries (162).[2] We are proud to serve the large international user community at SLAC, whose time on site can range from days to weeks to years. .

SLAC's expertise in the acceleration of electrons, in theoretical physics, and in the design and construction of particle detectors enables its researchers to pursue answers to basic questions about the structure of matter and about the fundamental forces that operate in our universe. To date, 3 Nobel Prizes in Physics have been awarded for research conducted here.[3] The laboratory's major high-energy physics program for the next decade is the B-Factory, which is a multi-national collaboration utilizing the PEP-II storage rings and the BaBar detector to investigate the asymmetry of matter and anti-matter.

At SLAC, scientists utilize the SPEAR synchrotron light source facility (being rebuilt in 2003 to become "SPEAR3") to probe the structure of matter at the atomic and molecular scale. The SPEAR facility is an electron storage ring that makes use of the intense, highly polarized x-rays emitted when electrons are forced to travel in a circular path. Such x-rays are highly prized by researchers in biology, chemistry, environmental and materials science, and related fields.

Although it began as a land-based, high-energy physics laboratory, research developments at SLAC and in both the high-energy physics and particle astrophysics fields have led to increasing collaboration among physicists, astrophysicists, and cosmologists to use advanced detector-based technologies in space. SLAC is presently collaborating on the Gamma Ray Large Area Space Telescope (GLAST) project, and this year has become the home of the new "Kavli Institute for Particle Astrophysics and Cosmology" which will foster collaboration between SLAC faculty and staff and Stanford faculty in the Physics and Applied Physics Departments.

SLAC has a continuing and important role in the development of new technologies, and maintains a deep and enduring commitment to the training of tomorrow's scientists and engineers. Educational programs at SLAC include graduate programs such as the SLAC Summer Institute, undergraduate science laboratory internships, a high school teacher curriculum development assistance program, an active tour program, a Visitor's Center, and a web-based "Virtual Visitor's Center."[4]

**History of SLAC Archives and History Office**

The SLAC Archives and History Office began its life in February 1986 as the "SLAC History Project." Bill Kirk, Assistant to the Director, and Louise Addis, Associate Head Librarian, began the project with a records survey in administrative groups throughout the lab. Identification of important records was followed by creation of an inventory database (SLACHIST) for some 500 separate records collections, and by the inauguration of a physical archive of important records no longer needed for current business. The records survey was followed up with an oral history program to gather information not fully documented in the available records.[5] A long-time SLAC employee, Marie LaBelle, with deep contacts in the SLAC community and wide knowledge of past projects at the site, was convinced to join the Project as Acting Archivist.[6]

Impetus for the SLAC project can be traced to several converging sources. The 1980's were marked by high interest in the history of particle physics both generally in the United States, and more locally at Stanford University. Early in the decade, the American Institute of Physics (AIP), working on contract with the US Department of Energy (DOE), completed a study of the records management and archives programs at several DOE contract laboratories. A final report and several guides for the selection and preservation of permanent records at physics laboratories resulted from this study.[7] Following the completion of their DOE project, AIP then initiated a much larger research project, called the "Study of Multi-Institutional Collaborations." To assist in organizing the new project it tapped – among others – Stanford Curator of University Archives Roxanne Nilan, known to the American Institute of Physics for her interest in the history of science and for her work on the AIP's Committee for the History of Physics.[8]

Joan Warnow of the AIP had been actively encouraging Bill Kirk and Louise Addis, as well as two successive SLAC Directors – W. K. H. Panofsky (1962-1984) and Burton Richter (1984-1999) – to take steps to preserve SLAC's history. Warnow also began encouraging Nilan to take an active interest in the history of SLAC, and to do what she could to support Kirk and Addis in their efforts. Locally at SLAC, awareness was growing among senior management that the laboratory was beginning an important transition period as the founding generation began to reach retirement age. Further motivation for the SLAC History Project was provided in 1982, when Peter Galison, Stanford University professor of philosophy and of physics, began conducting research on problems in the history of physics at Stanford, including the history of physics at SLAC.[9]

SLAC's History Project officially became the "SLAC Archives and History Office" (AHO) in Fall 1989, when Roxanne Nilan joined SLAC for a year's sabbatical to establish the new office to "evaluate, gather and make available" SLAC historical materials.[10] Nilan also continued to work as SLAC's and Stanford's representative on the AIP multi-institutional collaboration study. She was succeeded as head of the Archives and History Office by Robin Chandler, who served as SLAC Archivist from 1990 to 1995.[11] Throughout this period, Nilan, Chandler, Addis and Kirk made significant contributions to the American Institute of Physics' first, high-energy physics phase of their multi-institutional collaborations research by conducting oral histories, collecting data for a sociological census study, and supporting Peter Galison's related research on the history of the discovery of the J/Psi particle at SLAC in 1974.[12] During this period a number of publications – including a volume entitled <u>Big Science</u> – focused on the evolution of SLAC over time.[13]

The period 1993 to 1995 saw some growth in the SLAC Archives program, although staff support fluctuated. The program began a second growth spurt in mid 1996, when I was hired as permanent full-time archivist and, later that same year, when I hired a permanent halftime archives assistant. Work on a dedicated 2400-cubic-foot capacity state-of-the-art archival storage area was completed in 1996, and an Archives Program Review Committee comprised of internal and external stakeholders was established in 1999 to advise SLAC management on the goals, policies, and activities of the Archives program.[14] In 2000, a processing grant was awarded by the American Institute of Physics to support the arrangement of the papers of Burton Richter, SLAC Director and Nobel Laureate. By the end of calendar year 2002, the Archives and History Office had collected and at least partially processed over 1600 cubic feet of SLAC historical records, and had accumulated a processing backlog of roughly equivalent proportions.

**Rules We Live By**

As a United States government contractor, SLAC is subject to its governing agency's records regulations. Originally a contractor for the US Atomic Energy Commission (AEC), SLAC now contracts with an AEC successor agency, the US Department of Energy (DOE). DOE is, in turn, bound by the regulations of the US National Archives and Records Administration (NARA). Under the Federal Records Act, 44 United States Code (chapters 21, 29 and 33), it is the authority and responsibility of the Archivist of the United States to determine the retention and disposition of Federal Records.[15] This determination is made in consultation with the creating Federal agency and its contractors, and is documented with mutually approved instruments called "records retention schedules."

After a multi-year period of study and negotiation, a work group composed of NARA appraisal archivists, DOE and DOE-contract-laboratory records managers and archivists produced a records retention schedule for the research and development records of the Department. DOE Schedule N1-434-96-9, United States Department Of Energy Research And Development Records Retention Schedule, was approved by DOE and NARA for implementation in 1998. This schedule, heavily influenced by the findings of the just-completed AIP Study of Multi-Institutional Collaborations, bases appraisal and disposition of research records on the importance of the projects or experiments which created them.[16]

Experiments or projects are divided by the schedule into three levels. Level I experiments are those which received national or international awards of distinction; involve the active participation of nationally or internationally prominent investigators; or conduct research which results in a significant improvement in public health, safety, or other vital national interests. The significant records of Level I experiments are scheduled for permanent retention. Level II experiments involve research that leads to the development of a "first of its kind" process or product; improve an existing process, product, or application, or have implications for future research. Significant records of Level II experiments are scheduled to be retained for 25 years. Level III experiments are those which do not fall into Level I or Level II; the records of these experiments or projects are scheduled for ten-years' retention.

Significant records of a project or experiment are those created and maintained by the leadership of the experiment or project, and the records of the administrators who oversee the facility where the project or experiment was housed.

**How We Operate**

Records management and archives at SLAC, although separately administered, are closely coordinated operations. SLAC's Records Management Office is located in the Business Services Division, where the largest volume of temporary-retention, short-term-value records is created. The Archives and History Office is located in the Research Division of SLAC, where most of the long-term retention, continuing-value records of the organization are created and used.

All operating units at SLAC are requested to designate a "Records Liaison" for their unit: this individual is the major point of contact for Records Management and the Archives and History Office, and receives periodic training from the laboratory Records Manager and Archivist on records policies and procedures. Day-to-day guidance for the handling of archival records is available by personal consultation and from the Archives and History Office website,[17] which features a separate section for Records Liaisons with instructions for records storage and transfer, links to records disposition schedules, and definitions of records and archives-specific terms.

Once records have been transferred to the Archives and History Office, they are processed according to a "triage" approach. Basic processing is accomplished as quickly as possible on all receipts, and a skeletal database record is created for each accession. As accessioned records are consulted in response to reference queries, they are processed further, and their electronic guides are improved and expanded. This second-level processing is sometimes followed – when resources permit and the importance of the records warrants – by the more traditional folder-level archival processing and collection guide preparation.

Images transferred to the Archives and History Office, including photographs and drawings, are handled individually, and are indexed in a separate database (PHOTOINDEX). Since 1999 this database has been web-accessible[18], and in 2002 thumbnail images were added to the database records for a subset of the "most-requested" images.

Reference requests come to the Archives and History Office via personal visits, telephone calls, correspondence, and email.

The Archives and History Office web site, created in 1997, consists of a suite of pages grouped by the topics resources, policies and procedures, and historical information (Figure 1). The "Resources" portion of the web site allows access to our PHOTOINDEX database, and to a database glossary of SLAC-related vocabulary and acronyms we maintain called "SLACSpeak." The "Policies and Procedures" pages provide guidance to SLAC records liaisons, and to researchers wishing to use our collections. The "Historical Information" portion of our site provides highlights and milestones of SLAC history, information about the Nobel Prizes won for research completed at SLAC, and responses to a series of SLAC history-related "frequently asked questions."

On a monthly basis, the Archives and History Office now averages 9 personal-contact reference requests, and 26,000 web site hits – 90 to 95% of which visit a particular feature of our site described below. Personal contact reference requests have dropped significantly since the time that thumbnails of the most requested photos have been made web-accessible.

**Challenges and Opportunities**

Challenges facing the SLAC Archives and History Office are both physical and intellectual. "Getting the goods," that is, getting materials deposited in the archives, fits both categories. Our large community of international users is a fluid population with varying sources of support, affiliations, and connections to SLAC. As such, they pose a unique challenge for the archival program. Many of them create records that belong to them personally or to their home institutions, but some of them create records that are appropriate for inclusion in the archives at SLAC. In sorting out what belongs where, we emphasize the importance of preserving significant records in the appropriate repository – whether at SLAC or at another institution. We work with records liaisons, individual researchers, collaboration committees, administrative associates, and sometimes the Site Engineering and Maintenance Department, to locate records; identify, appraise and collect abandoned records; and explain to all relevant parties what records should be retained and when they should be retired.

Another physical challenge is the size and nature of our processing backlog. Although the laboratory has been around for 40 years, the Archives and History Office has been in operation less than half that time, and has some serious catching up to do. Some of the backlog materials have been at least viewed by current staff, but many of them are and will remain "mystery boxes" until time and resources can be found to complete the most basic level of processing on them. A special FY 2003 backlog processing project, funded by the SLAC Research Division, will assist us in solving some of the mysteries.

One of the biggest physical challenges facing our operation is the lack of storage space on site at SLAC. We have nearly reached capacity for our archival storage area, and in the past year have had to move all remotely stored backlog materials to an offsite location as the need for laboratory, shop and office space on site has grown.

The most significant intellectual challenge we face is the one posed by electronic records. SLAC has been at the leading edge of some developments in computing in physics, and has been an early adopter in others. We have a large backlog of experimental data tapes, as well as volumes of new materials that have been born digital. Like other archives around the world, we are struggling to find the most appropriate methods to identify electronic records of continuing value and to preserve them so that they are useful – and useable – in the long term. While the computer scientists working with the BaBar experiment wrestle with what is arguably the largest database in the world (as of Thu Apr 3 00:01:10 2003, over 751.5 TB had been stored in 580,850 files), the archives must plan to deal with an equally intimidating constellation of BaBar collaboration electronic technical notes, newsletters, email messages, design drawings, and specifications.

A small pilot project undertaken in the electronic records area has been the documentation of SLAC's early web site: the first one in the United States. Working with the SLAC "Web Wizards" who developed and maintained the site, and with special support from the Research Division, the Archives and History Office has been able to document the development of the first pages and the first site at SLAC. We have collected both paper and electronic records of the site, and were able to mount an online exhibition on our early web in time for its 10[th] anniversary on 13 December 2001.[19] These particular pages – this online exhibition – are the features of our web site that receive 90-95% of the monthly traffic (Figure 2).

**Conclusions**

Ongoing experiments at SLAC present the Archives and History Office with an opportunity to develop a plan for electronic archiving that collects records as they are created, but they also present a challenge, given that there are currently no storage media standards nor any well-developed tools for electronic archiving. While keeping a keen eye on electronic records archiving developments abroad and in the US, we are beginning work on developing a protocol to archive the BaBar experiment's electronic records by developing a digital equivalent of collecting and accessioning boxes of documents as they are created.

Tightening government budgets for scientific research impact all levels of laboratory operations, including archival efforts. In the US there have been encouraging signs that support for high-energy physics research is on the upswing, but those signs have not yet translated into improved budget totals. For this reason, flexibility in meeting the needs of the SLAC community, and in meeting the requirements of our government oversight agencies, will continue to be an important job requirement in the Archives and History Office.

However, despite recent funding challenges, the level of resources provided by SLAC to the archival efforts is at an all-time high, and over the past few years the Archives and History Office has matured into a program that is serving the needs of the SLAC community as well as preserving the history of the important scientific work performed at SLAC.

───────────────────────────────────────────

Acknowledgements

I wish to thank Peter Harper, National Cataloging Unit for the Archives of Contemporary Science (UK) for the invitation to present this talk to the CASE Conference. Thanks are also extended to Louise Addis, Joe Anderson , Robin Chandler, Anita Hollier, Pat Kreitz, Roxanne Nilan, Laura O'Hara, James Reed, and SLAC's Library and its Communications Group for their review of and comments on the work in progress.

This work was supported by the US Department of Energy Contract No. DE-AC03-76F00515.

───────────────────────────────────────────

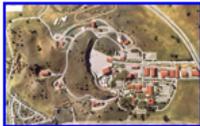

Figure 1: SLAC Archives and History Office web site home page
(http://www.slac.stanford.edu/history)

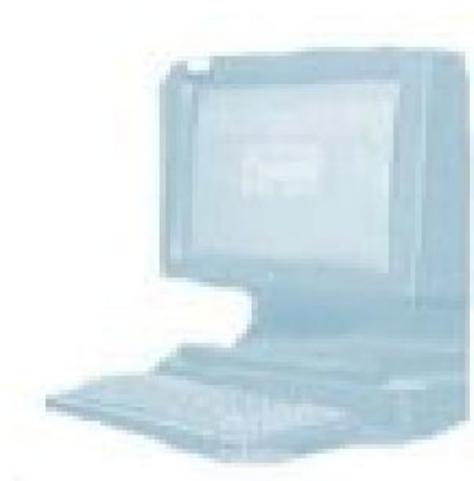

Figure 2: Home page of the SLAC Archives and History Office
online exhibit on the early web at SLAC
(http://www.slac.stanford.edu/history/earlyweb/)

REFERENCES and BIBLIOGRAPHY:

Symposium on Particle Physics in the 1950s, Batavia, Ill., May 1-4, 1985). New York: Cambridge University Press, 1989.

The Rise of the standard model: Particle physics in the 1960s and 1970s: Proceedings. Edited by Lillian Hoddeson, Laurie Brown, Michael Riordan, and Max Dresden. (International Symposium on the History of Particle Physics: The Rise of the Standard Model, 3rd, Stanford, Calif., 24-27 Jun 1992). New York: Cambridge University Press, 1997.

United States Department Of Energy. United States Department Of Energy Research And Development Records Retention Schedule (N1-434-96-9) 1998. (http://www-it.hr.doe.gov/records/doe_rd2.htm, 21 March 2003)

United States National Archives and Records Administration (NARA). NARA Basic Laws & Authorities. General Counsel and Policy and Communications Staff, National Archives and Records Administration. 2000 Edition. (http://www.archives.gov/about_us/basic_laws_and_authorities/basic_laws_and_authorities.html, 21 March 2003)

Warnow, Joan et al. A Study of Preservation of Documents at Department of Energy Laboratories. New York: American Institute of Physics, 1982.

Wolff, Jane. Files Maintenance and Records Disposition: A Handbook for Secretaries at Department of Energy Contract Laboratories. (DOE Report No. C00-5075.A000-16) New York: American Institute of Physics, 1982, Revised 1985.

NOTES:

[1] Neal (1968).
[2] http://www.slac.stanford.edu/slac/media-info/glance.html (3 April 2003).
[3] 1976: Burton Richter (shared with Sam C.C. Ting) for the discovery of the J/Psi particle. 1990: Jerome Friedman, Henry Kendall and Richard Taylor "for their pioneering work in the discovery of a heavy elementary particle of a new kind." 1995: Martin Perl (shared with Frederick Reines) for discovery of the Tau Lepton.
[4] Material in this section, as well as additional information on research and educational programs, is available from http://www.slac.stanford.edu/welcome/aboutslac.html (20 March 2003).
[5] Addis and Kirk (1987) and "Historical Chronology" section of Chandler (1995). Date and volume data are from SPIRES database SLACHIST (21 March 2003).
[6] Per email communication, L. Addis to J. Deken, 4 April 2003.
[7] Guidelines for Records Appraisal... (1982); Warnow, et al. (1982); Wolff (1982).
[8] Nilan was co-founder, along with Henry Lowood, of the Stanford University Libraries' "Stanford and the Silicon Valley Project," documenting the rise of microelectronics and personal computing in Northern California as well as the evolution of academic science and technology on the campus.
[9] Meanwhile, Fermi National Accelerator Laboratory (Fermilab) in Illinois was sponsoring a series of international symposia on the history of particle physics. The first two, *The Birth of Particle Physics* (1980) and *Pions to Quarks* (1985), had been held at Fermilab; the third was co-sponsored by SLAC and Fermilab, and held at SLAC on June 24-27, 1992. Participants in the third symposium, *The Rise of the Standard Model: Particle physics in the 1960's and 1970's*, included five SLAC staff members.

[10] R. Nilan, undated essay, 02-026, box 1; Chandler (1995); Stanford Historical Society *Sandstone & Tile,* Summer 1990 p. 12.

[11] Roxanne Nilan was with the AHO from 1989–1991 (full-time 1989-1990). Robin Chandler was on the AHO staff from 1990-1995 (full-time, 1990-1993).

[12] Chandler (1995) and (1991).

[13] Galison and Hevly (1992)

[14] The program review committee meets biennially, and its reports are available online at http://www.slac.stanford.edu/history/progrev/charge.html.

[15] NARA (2000)

[16] See "Review Philosophy and Guidelines" section of N1-434-96-9.

[17] http://www.slac.stanford.edu/history/

[18] http://www.slac.stanford.edu/history/photos.shtml

[19] SLAC Special Collection., World Wide Web 00-072. Exhibit is online at: http://www.slac.stanford.edu/history/earlyweb/index.shtml